# Quantum *K*-medians Algorithm Using Parallel Euclidean Distance Estimator

Amanuel T. Getachew

December 20, 2018

Department of Information Technology, Wolkite University, Ethiopia,
amanuel.tamirat@wku.edu.et , +251986521248

*Abstract*

*Quantum machine learning, though in its initial stage, has demonstrated its potential to speed up some of the costly machine learning calculations when compared to the existing classical approaches. Among the challenging subroutines, computing distance between with the large and high-dimensional data sets by the classical k-medians clustering algorithm is one of them. To tackle this challenge, this paper proposes an efficient quantum k-medians clustering algorithm using the powerful quantum Euclidean estimator algorithm. The proposed quantum k-medians algorithm has provided an exponential speed up as compared to the classical version of it. If and only if we allow the input and the output vectors to be quantum states. It shown that the quantum subroutine that computes the median from a set improves the time complexity from $O(m^2)$ to $O(m\ log\ m)$, where m is the size of the set. The proposed algorithm implementation handled in python with the help of third-party module known as QISKit. The implemented quantum algorithm was executed on the IBM Quantum simulators through cloud. The results from the experiment and simulation suggest that quantum distance estimator algorithms could give benefits for other distance-based machine learning algorithms like k-nearest neighbor classification, support vector machine, hierarchical clustering and k-means clustering. This work sheds light on the bright future of the age of big data making use of exponential speed up provided by quantum theory.*

## 1 Introduction

Quantum computers are in principle able to solve hard computational problems much more efficiently than classical computers hence it has attracted many interests. Quantum computation is an entirely new way of computing. It concerns the investigation of computational power by the implications of quantum mechanics for information processing purposes (Nielsen and Chuang 2010). The basic unit of information in today's digital computer known as a *bit*. In analogue manner, the basic unit of information in a quantum computer is called a quantum bit or *qubit* for short. A classical bit is restricted to hold the value (being in the state) 0 or 1. On the other hand, a qubit is not restricted to its analogous states it can be on both states $|0\rangle$ and $|1\rangle$ simultaneously, this kind of state is known as a *superposition* state. Quantum machine learning, an emerging field of research area, has demonstrated its potential to speed up some of the costly machine learning calculations as compared to the existing classical approaches. Since unsupervised machine learning problems involves tedious computations on massive data, the process is computationally intensive this is a good reason to look at how quantum computers accelerate the processes of k-medians clustering algorithm.

K-medians algorithm is the cluster analysis technique. It is used to group data instances based on the basis of their pairwise distances into $K$ partitions. The algorithm starts by specifying the number of clusters required i.e. $K$ in advance. Then $K$ points from the dataset are chosen randomly as cluster centers. Then all instances are assigned to their closest cluster center according to any distance function such as Euclidean or the cosine dissimilarity. Next $K$ new centroids or medians are recalculated from each of the clusters. These centroids are taken as center values for their respective clusters. Finally, the whole process is repeated with the updated cluster centers. Iteration continues until the same centroids are computed in the consecutive rounds. However, k-medians may not converge: it may oscillate between medians. Moreover, K-medians does not actually need the data points, given only the distances between dataset the algorithm works fine (Wittek 2014).

The quantum version of *k*-medians algorithm has been designed by Aïmeur, Brassard, and Gambs (Aïmeur, Brassard, and Gambs 2013), based on the quantum minimization algorithm a.k.a. DH algorithm, and the model commonly used in the quantum information processing task known as quantum Blackbox query. This quantum version of *k*-medians algorithm provides a quadratic speed up as compared to the classical version of it. However, their approach assumes the sum of distances from a particular element to the average of the set must be given in advance. The algorithm uses two layered subroutines. Firstly, with the help of a given Blackbox, the Blackbox is assumed to accept a point or a vector and returns the sum of the absolute distance between the given vector and the mean of the vectors on the cluster. Secondly, the quantum algorithm for finding minimum is applied, in order to find the smallest value of this distance function. In general, this quantum algorithm used to select the median for the clusters quantumly.

Quantum computations help to overcome the main bottleneck of the distance-based machine learning algorithms by calculating distances between the vectors in high dimensional space.

Aïmeur, Brassard and Gambs (Aïmeur, Brassard, and Gambs 2006) introduces the idea of, cosine similarity measurement by using fidelity function, $Fid(|u\rangle,|v\rangle) = |\langle u|v\rangle|^2$, between two quantum states $|u\rangle$ and $|v\rangle$ (the cosine of the smallest angle between two vectors). The fidelity can be obtained through a simple quantum routine sometimes referred to as a *Swap Test* (Fig 1). Given the quantum states $|u\rangle$, $|v\rangle$ and $|0\rangle$ i.e. two states and the auxiliary control qubit which initially set to 0, the subroutine Swap Test expresses the overlap of two states in terms of measurement probability of control qubit being in state $|0\rangle$. If the probability $P(|0\rangle) = 0.5$ means that the states $|u\rangle$ and $|v\rangle$ are orthogonal, whereas the probability $P(|0\rangle) = 1$ indicates that the states are identical.

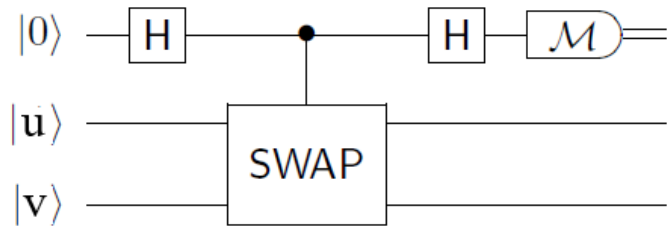

Fig 1: Circuit of the swap test circuit

Depending on the swap test ciruit, S. Lloyd, M. Mohseni, and P. Rebentrost (Lloyd, Mohseni, and Rebentrost 2013) proposed a quantum algorithm that can estimate the Euclidean distance between a given point and the set of points, by taking the average of the set. The classical version of this algorithm requires $O(Md)$, where $M$ is the number of elements found in the set and d is the dimension of the vectors but this quantum algorithm estimates the distance in $O(\varepsilon^{-1} \log_2(dM))$, time complexity. Moreover (Lloyd et al. 2013) shows the experimental implementation of a quantum k-means algorithm that has been done via a quantum computational model known as adiabatic (D-Wave). Recently, the quantum k-means algorithm using the cosine dissimilarity (Zhou et al. 2017) and the quantum hierarchical clustering algorithm based on quantum Euclidean estimator (Kong, Lai, and Xiong 2017) has been implemented. Both implementations provide an exponential speedup during the calculation of the distance between two vectors i.e. their exponential factor is determining by the dimension of the vectors only.

Now let us see how can we estimate the distance between a single point and the set of points. Having $2^n$-dimensional vector $\vec{u}$ and the cluster $V$ given by V = $\{\vec{v}_0, \vec{v}_1, \vec{v}_2, \dots \vec{v}_{M-1}\}$ the algorithm can estimate the distance between $\left|\vec{u} - \frac{1}{M}\sum_{j=1}^{M} \vec{v}_j\right|$. For the sake simplicity we assume $M$ is a power of two value. Then it is sufficient to have $\log_2 M$ qubits. In addition, we require one ancilla qubit. The first step is to prepare the state $|\psi_0\rangle$ in the form:

$$|\psi_0\rangle = \frac{1}{\sqrt{2}}\left(|0\rangle + \frac{1}{\sqrt{M}}\sum_{j=1}^{M}|j\rangle\right) \qquad (1)$$

The state $|\psi_0\rangle$ can be constructed by applying a unitary $V$ gate on $\log_2(M+1)$ qubits, the unitary $V$ operation can be constructed by using Toffoli, hadamard, phase-shift and NOT gates, it should be such that (Wiebe, Kapoor, and Svore 2014):

$$|V_{j0}| = \begin{cases} \frac{1}{\sqrt{2}}, & if\ j = 0; \\ \frac{1}{\sqrt{2M}} & otherwise. \end{cases} \qquad (2)$$

On the second step we are going to simulate and use the quantum random access memory (QRAM), which given as an oracle $O_v$ and able to access the classical vector information in the form of quantum state (the normalized vectors and their norms). The oracle can be simulated by the controlled – rotation around y-axis by the specified angle in the vectors (Zhaokai et al. 2014). This QRAM assumed to store the input vector $|u\rangle$ in its first address and the set $V$ at j+1 index address:

$$O_V\ |j\rangle|0\rangle = |j\rangle|V_j\rangle \qquad (3)$$

Querying $O_V$ by $|\psi_0\rangle$ results the state $|\psi\rangle = O_V\ |\psi_0\rangle|0^{\otimes n}\rangle$:

$$|\psi\rangle = \frac{1}{\sqrt{2}}\left(|0\rangle|u\rangle + \frac{1}{\sqrt{M}}\sum_{j=1}^{M}|j\rangle|V_j\rangle\right) \qquad (4)$$

Finally, via swap test circuit we project the state $|\phi\rangle$ on the state $|\psi\rangle$, the state of $|\phi\rangle$ is given as:

$$|\phi\rangle = \frac{1}{\sqrt{|\vec{u}|^2 + \frac{1}{M}\sum_{j=1}^{M}|\vec{v}_j|^2}}\left(|\vec{u}||0\rangle - \frac{1}{\sqrt{M}}\sum_{j=1}^{M}|\vec{v}_j||j\rangle\right) \qquad (5)$$

The state $|\phi\rangle$ can be generated by using quantum random access memory i.e. the previous defined oracle $(O_v)$, by querying the norms together with quantum simulation and apply the unitary transformation $e^{-iHt}$, where $H = \left(\sum_{j=0}^{M}|\vec{v}_j|\,|j\rangle\langle j|\right) \otimes \sigma_x$, to the state $\frac{1}{\sqrt{2}}\left(|0\rangle - \frac{1}{\sqrt{M}}\sum_{j=1}^{M}|j\rangle\right) \otimes |0\rangle$. Measuring the ancilla bit then yields the desired state $|\phi\rangle$ (Oshurko 2016).

Repeat the swap-test using the similar inputs several times to determine the probability of success of swap test P(0) up to an accuracy ε. We can find the classical distance from P(0) by:

$$\left|\vec{u} - \frac{1}{M}\sum_{j=1}^{M}\vec{v}_j\right| = \sqrt{2\left(|\vec{u}|^2 + \frac{1}{M}\sum_{j=1}^{M}|\vec{v}_j|^2\right)(2pr(0) - 1)} \qquad (6)$$

The following circuit (Fig *2*) summarizes the steps involved in the algorithm.

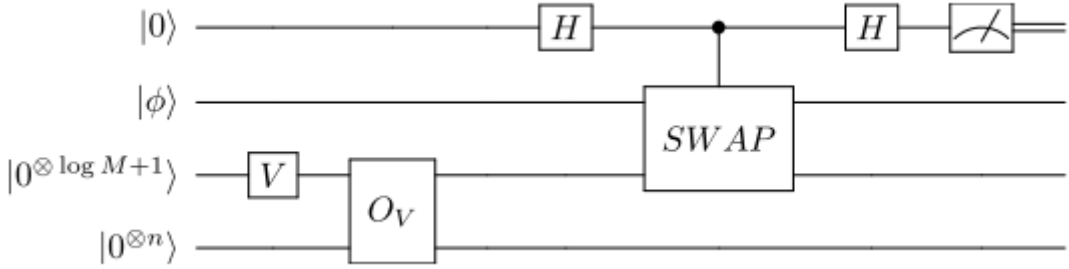

Fig 2: Quantum circuit for the Euclidean distance estimator algorithm.

## 2 Material and Methods

In recent years, quantum engineering technology, including preparation, manipulation, and detection of quantum systems, is undergoing rapid progress. The superconducting group in IBM has made first steps in 2016 that enables IBM to provide the first commercial quantum computing service called *IBM Quantum Experience* based on the circuit model approach. It includes actual quantum processors, a quantum software platform including a composer, application programming interface (API), compiler and tools, a quantum simulator and the Web application. It is an infrastructure as a service cloud based access to its experimental quantum computing platform (IBM 2016). Quantum algorithms can be designed by means of the QASM language and experiments can be executed by means of the QISKit Python SDK.

In this paper an efficient k-medians algorithm based on the above quantum Euclidean distance estimator algorithm (Lloyd et al. 2013) is proposed. The underlined quantum algorithm implementation is written in python, version 3.7 with the help of IBM Q Experience. For a successful demonstration of a quantum algorithm, first it is required to implement core circuits. So, a construction of many Controlled-U gates is fundamental. In this work up to 5 Controlled-U operations has been implemented by using a method taken from (Barenco et al. 1995). The generalized Toffoli-5 gate together with controlled-hadamard and Pauli NOT gates has been used to construct the state defined in Equation (1), and the controlled rotation around y-axis by angle θ has been used to construct the QRAM oracle that defined in Equation (3). The state of Equation (5) is simulated by the means of preparing an array of complex numbers which can be used as an amplitude for the required quantum register state. Then by using the initialize function, which is provided by QISKit, we have populated the amplitude to the given qubits.

The implemented quantum distance estimator algorithm utilizes four quantum registers and one classical register. The first quantum register is the ancilla qubit for controlling the swap gate, the second and third quantum register contains $(\log_2(M)) + 1$ and $(\log_2(d))$ qubits respectively in order to store Equation (4). Similarly, to represent the state of Equation (5) there are $(\log_2 M) + 1$ qubits in the fourth quantum register, where M is the size of the set and d is the dimension of each vector. The classical register contains only one classical bit, which used to store the value of the ancilla qubit during measurement. This quantum algorithm helps to speed up the process of computing a median from a set as shown in Algorithm **1**. The subroutine plays a great role by finding a median for each cluster in the k-medians algorithm.

---

1. *V = Accept a set of N dimensional points of size M*
2. *Min_dist = Quan_Parallel_dist(V[0], V)*
3. *Median = 0*
4. *for I = 1 up to M*
5.     *dist = Quan_Parallel_dist(V[i], V))*
6.     *If dist < Min_dist:*
7.         *Min_dist = dist*
8.         *Median = i*
9. *return Median*

*Algorithm 1: The proposed median calc algorithm*

---

# 3  Result and Discussion

This section discusses about the results found during the examination of the implemented algorithm. In IBM Q Experience the number of gates to be executed and the number of qubits involved in a particular algorithm are limited. Due to this, our dataset consists a few numbers of instances. A total of five experiments were performed. The first two experiments include a single vector and a set of vectors to be evaluated by the quantum distance estimator algorithm. The next experiment deals about the result found in the excution of the hybrid quantum median_calc algorithm. In the last two experiments, the quantum k-medians algorithm result are discussed.

In the first experiment we set a single vector, U, to be [2, 2] and the set V= [[2, 2], [2, 2]]. In similar manner for the second experiment U is [1.3, 0.9, 0.2, 0.6] and V =[ [0.5, 0.1, 0.7, 1.2], [0.6, 0.1, 0.2, 1.4], [1.4, 0.4, 0.5, 0.8], [0.2, 1.2, 0.8, 0.4], [0.1, 0.5, 1.7, 0.5], [1.3, 0.6, 1.2, 0.4], [1.1, 0.4, 0.5, 0.4], [1.2, 1.2, 1.0, 0.8]]. The results of the first two experiments are presented in histogram/bar graph representation. It is used to show the results of quantum algorithms since it is simple to understand. The height of the bar represents the fraction of instances the outcome occurs during the experiment. The highest bar in the graph is considered as the most frequent outcome as compare to the other and the bars which are too small for visualization are also the outcome of the experiment but with low probability of occurrence.

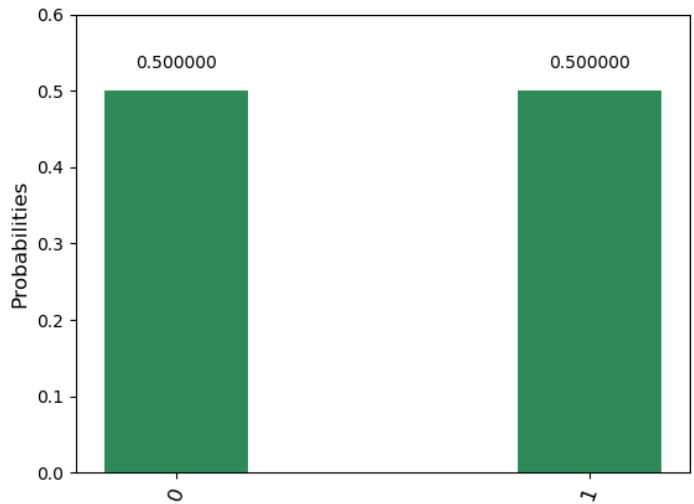

Fig 3: The probability of the ancilla qubit in the experiment one is 50% this means that the distance between the vector and the average of the set of vectors is zero.

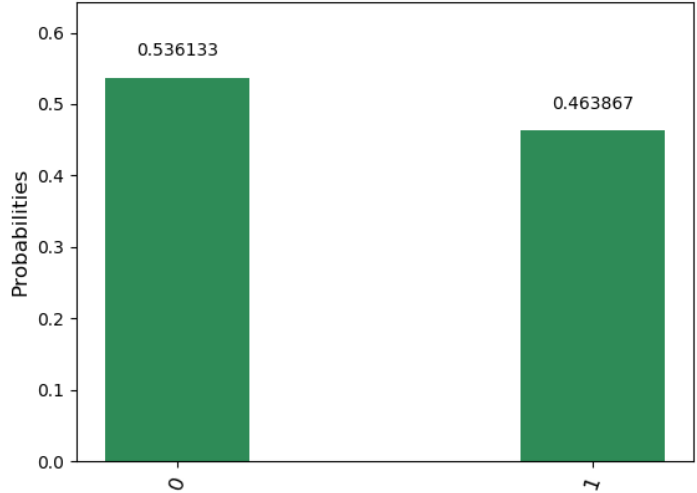

Fig 4: The probability of the ancilla qubit in the experiment two is 0.53613. That means, the algorithm estimates the distance between [1.3, 0.9, 0.2, 0.6] and [0.80, 0.5625, 0.8250, 0.73745] is around 0.9155 as compared to the real distance value 0.8794 it is a good estimation.

The third experiment demonstrates the result of a quantum median_calc algorithm. The algorithm was implemented to accept multi-dimensional array and returns the median index from the set. Technically speaking, the implemented quantum medians algorithm is a hybrid (quantum/classical) algorithm. The classical part of the algorithm iterates through the set and provides the i$^{th}$ value to the quantum part of the algorithm. The following table (presents results and variable values for the experiment. The median can be easily identified by selecting the minimum value from the distance results. The set that provided to the algorithm was: V= [[0.5, 0.1], [0.6, 1.4], [1.4, 0.4], [0.2, 0.4], [0.1, 0.5], [1.3, 0.4], [1.1, 0.4], [1.2, 0.8]]. The average value of this set is [0.8000, 0.54999]. As we can see the median of the given set is the 7$^{th}$ element i.e. when I equal to six.

Table 1: Results of quantum Median_calc algorithm.

| I | Vi | Z | Probability of zero | Quantum distance |
|---|---|---|---|---|
| 0 | [0.5, 0.1] | 1.5675 | 0.55078125 | 0.56426805 |
| 1 | [0.6, 1.4] | 3.62749 | 0.548828125 | 0.841722100 |
| 2 | [1.4, 0.4] | 3.42749 | 0.533203125 | 0.67469611 |
| 3 | [0.2, 0.4] | 1.5074999 | 0.545898435 | 0.526087044247 |
| 4 | [0.1, 0.5] | 1.5675 | 0.5751953125 | 0.6866400872 |
| 5 | [1.3, 0.4] | 3.157499 | 0.5244140625 | 0.55529236 |
| 6 | [1.1, 0.4] | 2.677499 | 0.509765625 | 0.323403530 |
| 7 | [1.2, 0.8] | 3.3875 | 0.517578125 | 0.488040565 |

The last experiment is used to evaluate the performance of the proposed quantum k-median algorithm to do this we have used eight two-dimensional vectors, v = [[2, 10], [2, 5], [8, 4], [5, 8], [7, 5], [6, 4], [1, 2], [4, 9]], to be cluster in three groups. The algorithm starts by choosing three random medians, [[6, 4], [8, 4], [7, 5]]. As we can see from Table 2, the algorithm took five iterations to complete the clustering task. In the first iteration, the clusters were formed based on the randomly chosen medians and the second iteration clusters was formed based on the newly calculated medians at the first iteration i.e. [[1, 2], [7, 5], [2, 10]].

In order to verify the effectiveness and feasibility of the proposed algorithm, we ran out method on IRIS dataset that contains 4-dimensional instances. The sample dataset has three class labels. We took three from two classes (first and third) and four instances from the second class. We brought 'precision' for better contrast effect which used to evaluate each cluster quality with respect to the number of instances from a particular class. The quality of clusters one and two scored 1, the third cluster contains two extra instances from class two, and the precision evaluation for this cluster was 0.8 see Table 3.

Table 2: Results of quantum k-medians algorithm.

| Iteration | Cluster 1 | Cluster 2 | Cluster 3 | New medians |
|---|---|---|---|---|
| 1 | [2, 5], [1, 2] | [7, 5], [6, 4] | [2, 10], [8, 4], [5, 8], [4, 9] | [[1, 2], [7, 5], [2, 10]] |
| 2 | [2, 5], [1, 2] | [8, 4], [5, 8], [7, 5], [6, 4], [4, 9] | [2, 10] | [[1, 2], [5, 8], [2, 10]] |
| 3 | [2, 5], [1, 2] | [2, 10], [8, 4], [5, 8], [7, 5], [6, 4] | [4, 9] | [[1, 2], [8, 4], [4, 9]] |
| 4 | [2, 5], [1, 2] | [8, 4], [5, 8], [7, 5], [6, 4] | [2, 10], [4, 9] | [[2, 5], [6, 4], [2, 10]] |
| 5 | [2, 5], [1, 2], [4, 9] | [8, 4], [7, 5], [6, 4] | [2, 10], [5, 8] | [[2, 5], [6, 4], [2, 10]] |

Table 3: Precision table of quantum k-medians algorithm. The overall cluster evaluation results (1+1+0.8)/3 = 0.93 or 93 %.

|  | Class 1 | Class 2 | Class 3 | n |
|---|---|---|---|---|
| **Cluster 1** | 3 | 0 | 0 | 3 |
| **Cluster 2** | 0 | 0 | 2 | 2 |
| **Cluster 3** | 0 | 4 | 1 | 5 |
| **c** | 3 | 4 | 3 | 10 |

## 4   Conclusion

Quantum machine learning, though in its initial stage, has demonstrated its potential to speed up some of the costly machine learning calculations when compared to the existing classical approaches. In this paper we have used the powerful parallel computing ability of a quantum computer to speed up the k-medians clustering algorithm. Finding a distance between a single point and the set of points is the core for realization of the algorithm. This parallel distance estimator algorithm uses quantum entanglement between the quantum state representation of the vectors and the ancilla qubit. Then a projective measurement is performed on the auxiliary particle alone in order to find the required distance value. The algorithm helps to easily find a median from a given set of vectors that used in quantum k-medians clustering algorithm.

In comparison to the previous quantum k-medians algorithm (Aïmeur, Brassard, and Gambs 2013), the proposed algorithm quantizes the subroutine that computes the median from a set in similar to the previous one. However, the proposed quantum k-medians algorithm improves the time complexity of computing the median from $O(m\sqrt{m})$ to $O(\frac{1}{\varepsilon}m \log m)$. Moreover, the implemented quantum k-medians algorithm does not require the sum of distance between each element is given in advance.

The proposed quantum k-medians clustering algorithm sketched in above runs in polynomial time $O(\varepsilon^{-1}I(Mk \log d + kM \log(Md)))$.where $M$ is the number of instances, $d$ is the dimension of the vectors, $I$ is the number of iteration and $k$ is the required number of clusters. This time complexity provides an exponential speed up as compared to the classical version of the algorithm which requires $O(I(Mkd + KM^2d))$.

This work also shows different experiments on the proposed quantum k-medians algorithm on a thirty-two qubits local and remote simulators. The experiments utilize elementary examples for testing the working performance of the algorithm. The results from the experiment and simulation suggest that quantum distance estimator algorithms could give benefits for other distance-based ML algorithms like k-nearest neighbor classification, support vector machine, hierarchical clustering and k-means clustering. This work sheds light on the bright future of the age of big data making use of exponential speed up provided by quantum theory.

# 5 Refences